# Mapping the Potential and Pitfalls of "Data Dividends" as a Means of Sharing the Profits of Artificial Intelligence

Nicholas Vincent, Yichun Li, Renee Zha, Brent Hecht
PSA Research Group, Northwestern University
nickvincent@u.northwestern.edu

**ABSTRACT**
Identifying strategies to more broadly distribute the economic winnings of AI technologies is a growing priority in HCI and other fields. One idea gaining prominence centers on "*data dividends*", or sharing the profits of AI technologies with the people who generated the data on which these technologies rely. Despite the rapidly growing discussion around data dividends – including backing by prominent politicians – there exists little guidance about how data dividends might be designed and little information about if they will work. In this paper, we begin the process of developing a concrete design space for data dividends. We additionally simulate the effects of a variety of important design decisions using well-known datasets and algorithms. We find that seemingly innocuous decisions can create counterproductive effects, e.g. severely concentrated dividends and demographic disparities. Overall, the outcomes we observe – both desirable and undesirable – highlight the need for dividend implementers to make design decisions cautiously.

**Author Keywords**
data dividends; data labor; machine learning

**INTRODUCTION**
There is a growing belief in economics that intelligent technologies are contributing to rising economic inequality [9, 10, 45, 63]. There is also substantial concern that this trend will accelerate as intelligent technologies become more capable [6, 16]. Critically, economic inequality is not an isolated phenomenon: it has been linked to political instability [3, 77], reduced national economic growth [69], financial crises [3, 8, 17], and even health outcomes [17], among other major societal challenges.

Given the exceptional stakes of large-scale economic inequality and computing's potential role in exacerbating this inequality, there have been growing calls in computing to much more seriously engage with our field's potential economic harms. Within the human-computer interaction community, mitigating computing's effect on economic inequality has been proposed as the "HCI problem of our time" [13, 27]. Within artificial intelligence, this topic is being cited as a top research challenge for the field (e.g. [2]).

In the many discussions about how to create a computing field that more broadly distributes its economic winnings, one idea has rapidly grown in prominence: giving people *data dividends*. While there is no commonly-agreed-upon definition of data dividends, the idea generally refers to giving users a share of the profits from an intelligent technology when their data contributions have been used to train that technology (e.g. [6, 27, 32, 42, 44, 50, 57, 73]). This approach is motivated by the fact that the intelligent technologies implicated in exacerbating inequality are highly dependent on data contributions from their users.

Payment for data has been well-studied in HCI through the lens of crowdwork. However, data dividends are both distinct from and complementary to existing crowdwork approaches. Key potential benefits of implementing data dividends as part of a broader solution set (that could incorporate crowdwork) include the ability to reward people for their implicit behavioral data and past data contributions. In addition, data dividends may help address concerns related to the working conditions of crowdworkers [24, 33, 64].

Interest in data dividends has grown enough that politicians are beginning to call for them: the governor of California proposed a state-wide data dividend in a major policy speech [72] and similar conversations are occurring in other jurisdictions [48]. However, despite their prominence, we have little understanding of the design space for data dividends and the challenging choices entities who want to implement data dividends must make. For instance, to which technologies would data dividends apply and to whom should dividends be given? Should the dividend be the same for everyone, or – as much of the discussion thus far suggests – should each contributor's dividend be "meritocratic", a function of the impact of that person's data on AI performance?

In concert with our limited understanding of the design space of data dividends, we also have very little information about the impacts of specific design choices. Critically, it could be that certain choices lead to data dividends that are concentrated among a small group, potentially even exacerbating inequality and contravening the purpose of data dividends. Given the HCI and social computing literature on the highly unequal nature of data contributions to many online social systems, it is reasonable to hypothesize that this antithetical outcome is a serious risk.

---

*This is a working draft. It has not been peer-reviewed and is intended for internal discussion in the computing community.

The goal of this paper is to start the process of developing an empirically-grounded decision framework for implementers of data dividends. We contribute to this process in two ways: (1) we *concretize* key aspects of the data dividends design space and (2) we explore this design space in a large series of *simulations*. With regard to the former, drawing from the ongoing discussions and our own experience implementing simulations, we find that the design space is more complex than early discussions suggest and highlight fruitful opportunities for future work. With regard to our simulations, we focus on simulating choices in the design space that are both *prominent in discussions* and *feasible to simulate*. We use multiple well-known machine learning (ML) approaches - logistic regression, support vector machines, and neural networks - and ten datasets that cover different domains and user-to-observation cardinalities (whether a user can contribute one or many observations).

At the highest level, our results emphasize that it will be absolutely critical to carefully weigh and simulate design decisions to ensure that data dividends will not do more harm than good. In particular, our findings raise concerns that the commonly-discussed "meritocratic" approach to data dividends could have unexpected and economically undesirable outcomes, e.g. dividends that are much more unequal than the United States income distribution. We also find that design choices that lead to highly concentrated dividends can also create serious disparities along demographic lines that correspond to legally protected classes, e.g. sex and race [71]. For instance, we observe many cases in which one demographic group receives less than 80% the median dividend of another.

Below, we begin with an overview of work that motivated our investigation, followed by an articulation of our design space research and our simulation research. We close by discussing implications, limitations, and future work. Before continuing, however, it is important to note that our goal is not to make an argument for whether the winnings of intelligent technologies *should* be distributed more widely through data dividends; for this, we refer readers to ongoing discussions in both the academic literature and popular media (e.g. [6, 27, 32, 42, 44, 50, 57, 73]). Our aim is to inform the growing data dividend discussion with concretization and empirical information from simulations. More generally, this paper is motivated by an argument that has been expressed strongly in the HCI community: the potentially tremendous stakes of computing's relationship to economic inequality demand that our field explore long-term plans to change this relationship [13, 27].

# RELATED WORK

## Background on Data Dividends

Critical to the stakes of the data dividends discussion is the growing support for data dividends from political leaders and the media. Most notably, the governor of California announced interest in implementing a data dividend in a major policy speech [72]. Similarly, many high profile media outlets have begun to cover data dividends and related concepts (e.g. [21, 56, 72, 80]).

There is also a growing body of literature that has advocated for data dividends and related approaches as a potential way to address computing-induced economic inequality. Posner and Weyl argued that providing payments for data, catalyzed by "data unions", could be a critical step towards addressing economic inequalities [57]. Arrieta Ibarra and colleagues introduced a theoretical framework that calls thinking of "data as labor" and rethinking the current paradigm in which data is treated as "exhaust" that companies collect; this framework directly supports the arguments underlying data dividends [6]. While data dividend discussions have recently grown in momentum, related ideas date back to the late 1990s, when a variety of micropayment schemes were proposed to pay for Internet content [61, 68]. Lanier has described a similar micropayment economy for the modern Internet [44].

Changing regulations around data also potentially pave the way for easier implementation of data dividends. Recently, the EU (e.g. GDPR) and California (e.g. CCPA) have begun to regulate data privacy [38, 46]. Legal standards around how data is handled and increased public attention to data's value may serve as stepping stones for data dividend-related policies.

## Relationship to Crowdwork Marketplaces

A substantial literature in HCI and related fields has emerged around crowdwork marketplaces such as Mechanical Turk (e.g. [4, 24, 29, 36, 37, 39]). While crowdwork marketplaces are well-studied, one reason data dividends have grown in prominence is that they complement the crowdwork approach in important ways. First, existing crowdwork marketplaces are not equipped to reward people for the data they produce during consumptive activities, i.e. "prosumption" [58]. Many profitable intelligent technologies (e.g. recommender systems and search engines) rely on prosumption for valuable behavioral data, e.g. what movies users watch and what links people click. Additionally, existing crowdwork marketplaces are transactional: workers are paid per contribution. This means existing crowdwork marketplaces are not well-suited to rewarding historical contributions, and thus cannot be used to share the profits of technologies invented in the years before any broad profit sharing approaches could be implemented. If computing-induced inequality increases dramatically, simply distributing earnings from future improvements to technologies will likely be insufficient to satiate economic and likely political demands. Finally, current crowdwork marketplaces have been subject to a broad range of critiques related to their suitability to addressing economic inequality, e.g. charges of excessively low wages [24] and terrible working conditions [33, 64]. By taking a much different approach, data dividends may be able to address some of these issues.

There is active work seeking to address some of the above issues, especially those related to wages and working conditions, and our expectation is that crowdwork could be an important part of broadly distributing the profits from intelligent technologies. While this paper focuses on non-crowdwork data dividends, below we further discuss how these ideas and literatures can benefit from each other.

**Data Valuation: What is Data Worth?**
A central challenge for data dividends is identifying the value of data. Our work intersects with this broad challenge at multiple levels, with two of particular note: (1) company-level data valuation, which is concerned with how much data is worth in aggregate to a company, and (2) observation-level data valuation, which is concerned with how much data is worth to a machine learning model.

Company-level data valuation is a subject of recent work [19, 23, 43, 47, 66], which has estimated the aggregate monetary value of data (e.g. using revenue from digital advertising). Notably, a study from Shapiro and Anejo called for 50% of estimated "data revenue" to be paid directly to Americans or into infrastructure [66]. While this recent company-level data valuation work provides a starting point for data dividends, future exploration will be needed. In particular, identifying the value of intelligent technologies not directly connected to revenue (e.g. services like AI assistants, search, image tagging, etc.) and public datasets on which for-profit intelligent technologies rely (e.g. Wikipedia [50, 74]) will be important for a more holistic implementation of data dividends.

As mentioned above, most discussions of data dividends have suggested (often implicitly) that dividends should be paid out in accordance with how much somebody individually impacted an intelligent technology. This "meritocratic" approach to data dividends requires some form of observation-level data valuation - a way to measure the value of data to a ML model – which is an open challenge that has been explored by recent ML research [22, 35, 40].

We focus on the *influence function* approach from Koh and Liang's work [40], which operationalizes meritocratic data valuation by estimating how model loss changes when an observation is removed. Using influence functions, introduced by Cook and Weisberg in the 1970s [14, 15], Koh and Liang showed that leave-one-out (LOO) loss change can be estimated *without retraining a model* and released open-source software to do so.

The ability to avoid retraining is absolutely critical for meritocratic data dividend simulations: retraining a model for every unit of training data – as is required for LOO evaluation – is computationally intractable, especially for complex models and large datasets. As such, influence functions were crucial to making our data dividend simulations computationally possible.

That said, while influence functions have rapidly come into wide use since Koh and Liang's work just two years ago, influence functions were evaluated by examining especially influential training data and misclassified test data. The data dividends context requires a different view, in particular one that considers all observations in a dataset.

To ensure that Koh and Liang's technique is sufficiently effective when adopting this broader view, we replicated and extended Koh and Liang's investigation of influence accuracy across entire datasets. As discussed in our simulation results section, while we found that influence functions were accurate for most data dividend needs, we also identified some potential weaknesses in influence functions. Despite this caveat, influence functions represent the only currently well-known approach for computationally feasible observation-level data valuation, which is required to simulate the prominently-discussed meritocratic approach to data dividends. We expand on potential extensions and alternatives to influence functions in Discussion.

**DESIGN SPACE**
Below, we first describe how we identified dimensions in the data dividend design space. Next, we describe each dimension and how we selected *prominent* and *feasible* choices within each dimension to explore via simulations.

**Identifying Dimensions of the Design Space**
A critical challenge in advancing the discussion about data dividends is that, despite the prominence of this discussion, the design space for data dividends has not been explored. To address this challenge, we followed the approach used by Quinn and Bederson in 2011 to map the design space of human computation [59], which entailed using existing literature to identify *dimensions* of the design space. While data dividends are not nearly as well-studied as human computation was in 2011, we were able to draw from scientific literature, ongoing discussions in the media, and conversations with a group of scholars interested in this issue to identify important design dimensions.

Communicating these dimensions – and the various choices that lie along them – requires the adoption of consistent terminology, which is also lacking in the nascent data dividends discussion. Here, we utilize the following nomenclature as it is well-suited at least to articulating our identified design dimensions: we refer to people who contribute *observations* (units of data) as *contributors*. Observations are organized in *datasets*, which are used to train models that perform some *task* (e.g. "classify spam emails") that plays a role in some intelligent technology. Very generally, a *data dividend implementation* is a scheme that takes a share of the economic winnings of one or more intelligent technologies and distributes these winnings to contributors in some manner.

Below, we introduce and describe the six dimensions we identified: *Funding, Implementer, Tasks, Time Period, Disbursement, and Observation Valuation*.

**1 – The *Funding* Dimension**
The *Funding* dimension considers how data dividends will be funded. One approach might be taxing revenue that is attributable to intelligent technologies. As highlighted in Related Work, some work on company-level data valuation has estimated the monetary value of data [19, 23, 43, 47, 66], but there are still challenges with such estimations. As such, alternatives might involve charging companies per unit of data they use (perhaps with price corresponding to the privacy implications of the data) or imposing a tax on "tech" companies (although company classification could prove challenging) [76].

In our simulations we do not assume any particular *Funding* design choice. To make our simulations feasible, we focus on analyzing results in terms of what fraction of the total funding each contributor would receive under any funding choice, i.e. the fractional distribution of dividends. This is possible because *Funding* is decoupled from the other dimensions – it only controls how the money for data dividends is collected, not how it is distributed.

**2 – The *Implementer* Dimension**
The *Implementer* dimension is concerned with which entity will actualize data dividends. One of the most prominent advocates for data dividends is the governor of California, suggesting government as a realistic implementer [72]. However, another possibility might be that tech companies themselves implement dividends. Alternatively, organizations analogous to labor unions could mediate interactions between data contributors and tech companies and thereby implement some form of data dividend [45].

Critically, different implementers will have different levels of *data access*. For instance, the California government may not be able to access every observation and model used by a tech company, and thus may be unable to implement certain types of data dividends.

In our simulations, we focus on an *Implementer* that has full data access, though we also consider simple approaches that do not require data access (such approaches are easier to simulate due to their necessary simplicity). Future work might further explore data dividends that only require partial data access.

**3 - The *Tasks* Dimension**
The *Tasks* dimension determines which tasks performed by intelligent technologies should be considered in implementing data dividends. The choice will determine for which datasets implementers must account. As data dividends discussions have included a variety of *Tasks,* we sought to cover a variety of task domains in our simulations, focusing on simple classification tasks and datasets to keep our early-stage simulations feasible.

Every dataset has a *contributor-to-observation cardinality*. Datasets that have only a single observation per contributor have a *one-to-one* cardinality. Any dataset in which each observation describes one person fits this category, e.g. a dataset describing medical or financial status of individuals. Many datasets have *one-to-many* cardinalities, e.g. a product review dataset in which users can review many products. Depending on the definition and scope of a single observation, it is possible to have *many-to-many* cardinality. For instance, a set of Wikipedia articles written by a group of editors could be seen as *many-to-many* if each article corresponds to a single observation. In our simulations we focus on the more commonly-available one-to-one and one-to-many datasets.

A critical concern relating to cardinality is that one-to-many or many-to-many cardinality datasets allow for large disparities in contribution, and thus dividends might reflect these disparities and could exacerbate inequality (because a small group of people get the majority of the dividends). For instance, the social computing literature has identified that a small group of people contribute a large percentage of the content (i.e. data) on Wikipedia, YouTube, and other social computing platforms [11, 25, 28, 53]). These contribution patterns may be a serious concern for data dividends, so we were careful to ensure our simulations included some one-to-many datasets.

**4 - The *Time Period* Dimension**
This dimension considers the *Time Period* in which observations will be considered for a data dividend. This seemingly minor consideration has enormous implications for both designers and beneficiaries of data dividends. If the time period covers the past, dividends will have the *backward-looking* property and can reward people for data that has already been contributed. The possibility of backward-looking data dividends is a key motivation for data dividends, as they can reward people for the data contributed in the years before broad distributing of economic winnings is implemented. The *Time Period* can also include the future, giving dividends the *forward-looking* property. If data dividends are forward-looking, they can reward future contributions and incentivize the collection of new data (similar to crowdwork).

Given the political importance of backward-looking dividends and the need for additional work on the effects of incentives, we focus on backward-looking *Time Periods* in our simulations. Below, we discuss how future work might further explore forward-looking dividends.

**5 - The *Disbursement* Dimension**
This dimension determines how the winnings of intelligent technologies will be disbursed. One way that different *Disbursement* choices can be categorized is by their granularity, ranging from extremely granular, e.g. every contributor receives a customized paycheck, to very low granularity, e.g. funds are pooled together and used for public projects like infrastructure or education [66]. Notably, low granularity approaches that do not attempt to customize dividends at an individual-level obviate the need for navigating the *Observation Valuation* dimension. As we will see, some of our results suggest this might be desirable. Beyond differences in granularity, *Disbursement* choices can also differ in other ways, e.g. which specific initiatives low granularity dividends might fund.

In our simulations, we assume individual-level *Disbursement* in which each contributor receives money, as this idea is most prominent in data dividend discussions.

## 6 - The *Observation Valuation* Dimension

The *Observation Valuation* dimension is concerned with measuring the value of observations to machine learning models, i.e. it addresses the observation-level data valuation challenge introduced in Related Work. This design dimension is highly complex in terms of the number of possibilities available to designers, the need to consider the technical details of machine learning models and, as we will see below, impact on the outcome of data dividends (and critically, influence whether data dividends will achieve their economic goals).

Individual-oriented "meritocratic" *Observation Valuation* has appeared prominently in discussions of data dividends [6, 57] and even machine learning literature [22, 35]. This approach entails assigning dividends based on estimates of the effect of individual observations on machine learning models. In theory, such estimation could involve a variety of evaluation procedures (e.g. LOO, Shapley value, leaving out an observation alongside other observations) and performance metrics (e.g. loss, accuracy, precision). One feasible meritocratic approach that has appeared in data dividend discussions involves examining how a machine learning implementation's "loss function" changes when leaving out individual observations. Loss functions are central to how machine learning algorithms operate: for a given machine learning implementation, a chosen loss function is the mathematical function that a machine learning algorithm seeks to minimize (lower loss means better performance). Given that influence functions make this approach much more feasible than alternatives, this is the approach we take in our simulations.

Another option for meritocratic evaluation is to sum the effects of all observations from each contributor, so that each contributor has a "summed influence". Influence can be positive or negative, so summed influence will not necessarily correlate with the number of observations contributed. As we will see in our results, this approach could mitigate serious concerns around contribution disparities leading to highly unequal data dividends.

Two approaches to *Observation Valuation* that do not require an *Implementer* with access to underlying machine learning models would be to simply assign the same dividend to each observation (such that a contributor's dividend is determined by their number of contributions) or to each contributor (such that observations beyond the first provide no additional value to a contributor). Variations might assign extra value to observations of a certain class that are externally visible (e.g. rare medical cases) or to contributors of an observable type (e.g. people living in rural areas). These straightforward approaches may be desirable, as they will have a low cost compared to approaches that require access to underlying models.

### Transforming Observation Value to Money

Individual-level *Disbursement* requires assigning each contributor a dividend, i.e. a fraction of the money determined by choices from the *Funding* dimension. This necessitates an open-ended and important *Observation Valuation* challenge: transforming value measurements (e.g. LOO loss change) into dividends (i.e. money). In theory, implementers can choose from a wide variety of mathematical functions to achieve this. In our work, we sought to explore the most straightforward of these functions with the goal of simulating how data dividends might be first implemented.

Specifically, we focus on four simple transforms that capture a variety of perspectives on data valuation. We assume that early data dividends would not allow for "data debt" and thus any transform must result in non-negative dividends (we expand on this in Limitations). The first three simple transforms use functions which operate on continuous values: *Shift*, *Absolute Value*, and *Clipping*. The fourth transform, *Binning*, treats measurements as ordinal. Keeping with our goal of exploring straightforward approaches, we ensure that each transform assigns fractional dividend values to each contributor that collectively sum to one (because each dividend represents a fraction of total funds). Below we discuss each of these four transforms in more detail.

*Shift Transform:* A basic approach to achieve non-negative dividends for all contributors would be using mathematical translation to shift the minimum measurement value to zero (i.e. subtract the minimum), and then scale the shifted values so that each contributor is assigned a fractional dividend value. This operationalizes a view in which an observation's monetary value is linearly related to its marginal effect on model performance.

*Absolute Value Transform:* In some cases, observations that are "hurtful" to a model may be useful to developers. As highlighted by Koh and Liang [40], "hurtful" points can help debug potential pitfalls like domain mismatch or wrongly-labeled data. One classic example of a constructive "hurtful" observation comes from the MNIST handwritten digits dataset: when classifying only "7" and "1" digits, a training observation "7" with an extra horizontal line (e.g.

"≠") will "harm" the model's ability to classify a "7" correctly (i.e. make it more likely the model classifies "7" as a "1) [40]. Developers can use this information to re-evaluate the validity of their model or data.

An absolute value approach can ensure that hurtful observations are rewarded alongside helpful observations. The basic implementation of this approach that we use involves taking the absolute value of each measurement and then following the *Shift* procedure.

*Clipping Transform:* This transform treats reductions in model performance harshly by giving all hurtful data contributions zero value and then scaling to fractional dividend values. This might make sense if the organization is already confident that they have a very representative test set (i.e. there are no ≠- like situations).

*Ordinal Binning Transform:* An alternate family of straightforward initial transforms considers ordinal value measurement and assigns certain rewards to certain categorical bins. For instance, one Binning approach might give all observations with loss change above or equal to the median loss change twice the reward of those below the median, so that very helpful observations receive greater rewards. Variants might assign rewards based on quintile, decile, etc. Critically, *Ordinal Binning* provides a simple way of controlling the outcome of dividends and imposing a maximum degree of inequality. As we will see below, this may be desirable to reduce the impact of *Observation Valuation* noise on the dividend a contributor might receive. To explore a basic and interpretable example of binning we use the "above median gets 2x" approach described above in our simulations.

## SIMULATING DATA DIVIDENDS

In this section, we describe the second major portion of work in this paper: our simulations of data dividend implementations. Whereas the design space in the previous section outlines a decision framework for dividend implementors, in this section we seek to provide empirically-informed guidance within that framework. As discussed above, we do not attempt to simulate all possible choices within the design space; doing so is not possible at this early stage of the literature. Instead, we sought to simulate choices that were both (1) particularly prominent in the growing discussion around data dividends and (2) feasible for simulation.

Using these criteria of prominence and feasibility, we focused our simulations on the dimensions of *Observation Valuation* and *Tasks*. As discussed in Related Work, *Observation Valuation* is a central topic in the current data dividends debate, with interest being particularly high in "meritocratic" approaches. Thus, we do a deep dive into the transform component of *Observation Valuation*, which speaks to how a meritocratic implementation might work. For context, we also include the much more straightforward approach that simply apportions dividends based on a fix amount per observation. The data dividend discussions also consider a wide variety of *Tasks* (and thus datasets). As such, we sought to include tasks from a wide variety of domains. Ensuring diversity of *Tasks* has the added benefit of a degree of self-replication, ensuring our results are not specific to a single domain. A summary of the choices we made within each dimension is shown in Table 1.

Our simulations all follow a similar structure. First, we implement an *Observation Valuation* approach. Next, we apply this approach across all of the *Tasks* we consider (which is made possible using influence functions). Finally, we examine for each approach and each task the effects on our core outcome metric: the equality of the resulting dividend distributions. With respect to this outcome metric, we consider both overall equality as well as, where possible, equality across demographic groups.

Below we first describe our methodological approach in more detail, including the required validation of influence functions for the full training data distribution use case (as discussed above). We then unpack the results of our simulations with respect to the equality of dividends.

### Methods

*Machine Learning Tasks and Modeling Approaches*

We simulated data dividends for ten different classification *Tasks* and their corresponding datasets. The *Tasks* we used include classification tasks related to finance (Adult Income [41], German Credit [30], Bank Marketing [52], and Boston Housing [26]), medicine (Heart Disease [18], Breast Cancer [70]), text (Spam [51], Yelp sentiment [60], Amazon sentiment [5, 49]) and images (Fashion MNIST [78]). Training dataset size ranged from 242 to 420k. We note that many of the datasets we used are hosted through the UCI Machine Learning Repository (UCI MLR) and scikit-learn [55], both of which have been widely used in research and practice, i.e. these datasets have played an important role in the long-term development of intelligent technologies.

Critically, the Yelp and Amazon tasks had one-to-many user-to-observation cardinality, so we were able to study the concerns around contribution disparities discussed above. Several datasets also included demographic data, which enabled our investigation of demographic disparities.

| Dimension | How Our Simulations Explore the Dimension |
|---|---|
| *Funding* | We make no assumptions about *Funding*. |
| *Implementer* | We assume the *Implementer* has data access. |
| *Tasks* | We include ten different tasks and three machine learning approaches. All tasks are simple classification tasks. |
| *Time Period* | We assume data dividends are strictly backward-looking. |
| *Disbursement* | We assume individual-level, "paycheck-style" *Disbursement*. |
| *Observation Valuation* | We consider multiple "meritocratic" approaches alongside a "fixed value per contributor" approach. We consider four different loss-to-dividend transformations that encode different understandings of data value. |

**Table 1. Shows the data dividend design dimensions we identified and their relationships to our simulations.**

In order to better capture how ML is used in practice, we performed our simulations using multiple modeling approaches: logistic regression (LR), support vector machines (SVM), and convolutional neural networks (CNN) [40] (the same set of approaches that that Koh and Liang used to demonstrate the utility of influence functions). As the Fashion MNIST task was created specifically to challenge CNNs, and CNNs are substantially more expensive to train than LR or SVM, we applied the CNN architecture from Koh and Liang to this one task and applied the LR and SVM implementations from Koh and Liang to our nine other tasks. Furthermore, because CNN training is expensive, Koh and Liang estimated the change in loss when training a model (such that loss approaches a local minima), removing an observation, and then retraining the model (for 30k steps), as opposed to a full retrain. We also adopt this approach. For each task, we randomly withheld 20% of the data to use as a test dataset, which we used to evaluate model test loss (for Fashion MNIST, we usd the test dataset provided by the dataset's creators).

*Data Dividend Outcome Metrics*
To compare the equality of data dividends implemented with different design choices, we focus on the Gini Index, a standard inequality metric from economics. The Gini index spans from 0 (total equality, everyone gets the same amount) to 100 (total inequality, one individual gets everything) [54]. While the exact impact of data dividends on overall inequality will depend on *Funding* choices and who receives data dividends, the Gini index allows us to compare how broadly each (simulated) implementation of data dividends shares the economic winnings of AI systems. To study the prevalence of demographic disparities, we look at the median dividend received by members of the demographic groups in our datasets.

*Validating Influence Functions for Data Dividends*
As noted above, a critical precursor to the evaluation of *Observation Valuation* design choices is the ability to accurately estimate the effect of each observation on model performance. Koh and Liang focused on evaluating how well influence functions could estimate the LOO loss change when considering the most influential training data and the test loss for a single misclassified test observation [40]. In these evaluations case, influence functions were very accurate, although slightly less so for SVM and CNN than for LR.

However, our use of influence functions requires evaluating them with a different perspective: we must consider the full spectrum of training observations, including those that are not particularly influential. As such, we replicated and extended Koh and Liang's accuracy evaluation for a random sample of 100 training observations per task and computed loss across a randomly sampled test set, rather than for a single misclassified test observation.

Overall, our experiments replicated Koh and Liang's finding that influence functions are accurate for estimating LOO loss, even for the broader data dividend use case. We observed that influence functions were most accurate for LR, moderately accurate for SVM, and less accurate for CNN. The average (across tasks) correlation was 0.93 for LR, 0.63 for SVM, and 0.24 for CNN (all individual correlations were statistically significant at $p < 0.05$). When we instead looked at the 50 most influential observations for the CNN model, the correlation was 0.79, further reinforcing the utility of influence functions for evaluating highly influential observations.

In general, the lower correlation for CNN we observed suggests that influence functions may have weaknesses in some data dividend contexts. Regardless, influence functions represent the only feasible approach to performing this type of *Observation Valuation* for expensive models like CNNs. *Implementers* likely will need to run their own validations (following a version of our approach) and make a judgment as to what level of accuracy is acceptable. This may lead some *Implementers* to seek alternatives or modifications to influence functions. We discuss such approaches below.

**Results**
In this section, we explore the outcomes of the prominent and feasible design choices that we explored.

*Inequality of Data Dividends*
Figure 1 shows the Gini index of data dividends that correspond to loss-based *Observation Valuation* for each combination of task, model and loss-to-dividend transform. Figure 1 also shows Gini index for a fixed amount per observation approach for one-to-many cardinality datasets, shown below the black horizontal line (doing so for the one-to-one cardinality datasets necessarily has a Gini index of zero). The Binning transform shown in Figure 1 is the implementation from Section 3: observations with above-median influence receive twice as much as below-median observations. Two baselines from real-world income distributions are also included as vertical dotted lines in Figure 1: the United States (red, at 39.1), Finland (blue, at 25.9), the country in the OECD, an intergovernmental economic organization, with the lowest Gini index [54].

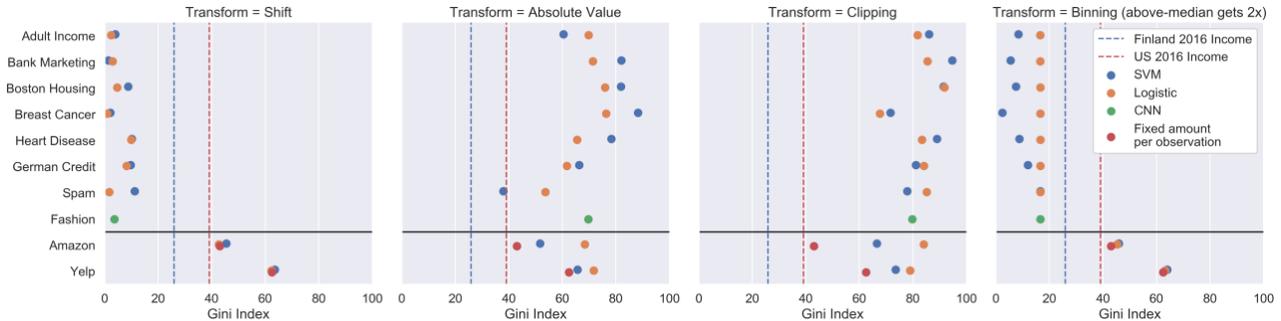

**Figure 1. Gini index for data dividends across tasks and models. A lower Gini index means more equality. Colors distinguish different models. Each row shows a different task. Dotted lines show country-level 2016 income Gini indices.**

Figure 1 displays several patterns that are highly informative regarding the distributions that arise from various choices within the *Observation Valuation* dimensions. First, focusing on the *Shift* transform (leftmost column) and the one-to-one tasks (above the black horizontal line), we see support for the notion that data dividends can result in highly equal distribution of winnings. Using *Shift*, across all tasks and models, we see low Gini index values, i.e. most contributors receive about the same dividend.

However, widening our view of Figure 1 to include the middle two columns, we see that with *Absolute Value* or *Clipping*, dividends become much more unequal, with some Gini index values around 80 (about twice the Gini index of the U.S. income distribution). As a representative example, Figure 2 shows the log-scaled histograms of the loss change distribution (left) and transformed dividend distributions (middle and right) for the Adult Income task and LR model. We converted the dividends in this figure from fractional values to dollars by assuming a total funding pool of $36k (the total number of contributors). This makes the mean dividend $1 and differences easier to interpret. With the *Shift* approach there are no dominant observations, i.e. "superstars" [62], but through transforms, one can induce superstars, e.g. the small group of contributors shown on the right earning 25-50x the average dividend.

Next, when looking at the one-to-many cardinality Yelp and Amazon tasks, we see high levels of inequality. In fact,

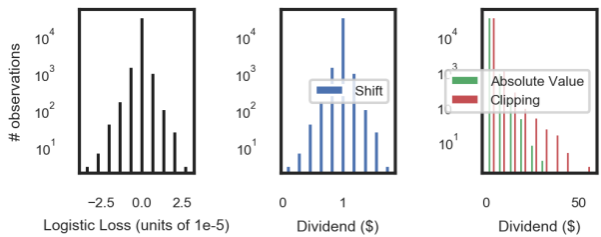

**Figure 2. Left: Histogram of loss changes for Adult Income dataset and logistic regression. Middle and Right: Corresponding dividends for *Shift*, *Absolute Value* and *Clipping* transforms, shown in dollars for a total *Funding* pool of $36k, the total number of contributors (i.e. so the mean dividend is $1). Counts are shown on a log scale.**

for these tasks, the level of inequality for the *Shift* transform is nearly identical to the level of inequality for "fixed amount per observation", i.e. contribution patterns are driving dividend disparities. *Absolute Value* or *Clipping* transforms make dividends more unequal. This suggests that loss-based *Observation Valuation* alone did not induce a diminishing margin return effect.

However, we did observe one design choice described above that could address these contribution disparities: summing loss change by contributor. While the results shown in Figure 1 use observation-specific value measurement, an implementer might instead sum loss changes per contributor and apply a measurement-to-dividend transform to this "summed loss". Critically, because loss changes can be positive or negative, the summed loss of each contributor does not necessarily correlate with their number of contributions. We simulated this dividend approach for the Amazon and Yelp datasets. The highest Gini index using this approach with the *Shift* transform was 1.6, i.e. this design choice resulted in very equal dividends.

Taken together, the variety of outcomes we observed highlights the importance of simulating various designs before implementation. A well-intentioned *Observation Valuation* choice like using an *Absolute Value* loss-to-dividend transform could lead to undesirable outcomes.

*Demographic Disparities in Data Dividends*
To investigate data dividends with a demographic lens, we used the simulated dividends from our three tasks with demographic features that correspond to protected classes. The Heart Disease and German Credit datasets use age and sex, and the Adult Income dataset uses age, sex, and race. We use the dividends from our LR models, for which influence functions are most accurate.

To explore the prevalence of demographic disparities in our data dividends, we looked at the median payment received by contributors within each demographic group. For age, we grouped contributors into "under 40" or "40+", based on how age is defined as a protected class in U.S. law [71]. For sex and race, we used the groups afforded by our datasets (male/female and U.S. Census racial categories). We computed "median dividend ratios" for each pairwise combinations within each set of groups, e.g. comparing the median dividend for "40+" and "under 40" contributors. We computed each ratio by dividing the smaller value of each pair by the larger value, so that each ratio was fractional (or one).

We investigated the prevalence of disparities with a heuristic: does one group have less than 80% the median dividend of another group? This heuristic was loosely inspired by the "80% rule" from U.S. Labor Law [12], used in the past to define *disparate impact* in terms of employer hiring decisions. Machine learning fairness researchers have adapted this rule to identify disparate impacts in terms of classification of false positives [20, 79].

We observed that for each set of demographic groups within each dataset, there was at least one transform for which one group received less than 80% the median dividend of another group. Moreover, the average disparities in median dividend were very large: the average median dividend ratio of medians was 0.24 for *Clipping*, 0.56 for *Absolute Value*, 0.7 for *Binning*, and 1.0 for *Shift*. In other words, every transform except *Shift* led to meaningful demographic disparities. For example, when applying the *Absolute Value* transform to the Adult Income task, the median dividend for a man was 2.5x that of a woman. For *Absolute Value* applied to the German Credit dataset, people under 40 received 1.5x the median dividend of those over 40.

## DISCUSSION

Our results indicate that the data dividend design space is much more complex than early discussions might suggest, and that certain design choices can lead to unintended outcomes. These findings mean it will be important to simulate the outcomes of any data dividend implementation before execution to avoid potentially counterproductive and harmful effects like increased inequality or disparate impacts across demographic lines. Put simply, data dividends need to be implemented with great caution if they are to achieve their stated goals. Below, we discuss some implications of this overall result.

### Customizing Outcomes of Data Dividends

While we approached our simulations from the standpoint of a data dividend *Implementer* comparing choices about how to value observations, one could also use these simulations to figure out how to induce a specific dividend distribution. In other words, an implementer could work backwards through the design space, by starting with a desirable *Disbursement* distribution they would like to create. In the framework of the growing fairness literature, this would be equivalent to allowing for disparate treatment to achieve a desired (non-disparate or less-disparate) impact [7]. Exploring these types of disparate treatment approaches in data dividends is a useful direction of future work.

Throughout our Results section, we discussed how highly unequal data dividend outcomes may be cause for serious concern, but also observed that there are a number of *Observation Valuation* choices that can lead to extremely equal dividends. Extremely equal outcomes may also concern some implementers, given that economists have raised concerns around enforced "equality of outcomes" (e.g. [31, 65, 75]). This means there may be tension around selecting an acceptable level of equality for data dividends.

It may be possible for data contributors to play a role in navigating the design space and providing input on an ideal level of inequality for data dividends. One way that such collaboration might come about is through collective bargaining, i.e. by data unions (or other types of MIDs described by Weyl and Lanier [45]).

### Contribution Disparities

Our results highlight that for one-to-many cardinality datasets, the long-tail contribution patterns in user-generated content that have been well-studied in social computing (e.g. [11, 25, 28, 53]) present major challenges to using data dividends to mitigate computing-induced inequality. This means that any attempt to pay people for their data must contest with a very uneven "observation contribution playing field". Our results suggest that the issue of contribution disparities, which so far has not been highlighted in the data dividend conversations, may need to be treated as a central concern.

To contend with the effect of contribution disparities, data dividend implementers may want to consider approaches such as introducing diminishing returns (i.e. taxation) or by using the "summed influence" approach we explored in our simulations (summing the LOO loss changes for each contributor). By simulating choices within the design space, implementers might create a menu of options which address these disparities.

Data markets may be another viable approach to addressing the issues with contributions disparities that we observed – providing financial incentives to contribute data might lead to a much larger population of heavy data contributors. However, data markets are only applicable to data dividend implementations with a forward-looking *Time Period* and to types of data which are compatible with a market approach (i.e. not implicit behavioral data).

### Beyond Influence Functions

As we discussed above, we were reliant on influence functions to make our simulations possible. However, there is ongoing research exploring alternate approaches. Recent work from Jia et al. [34, 35] and Ghorbani and Zou [22] used the *Shapley value* [67] for data valuation. The Shapley value is the average marginal effect an observation has on some outcome (e.g. loss) across all possible combinations of data, i.e. the Shapley value accounts for every possible "arrival order" of observations. However, in Jia et al.'s experiments, influence function-estimated LOO loss change was highly correlated with Shapley value and Shapley values were still much more expensive to compute than influence [35]. Future work could explore more efficient Shapley-based data dividends, perhaps extending techniques from Jia et al. [34].

*Implementers* might also seek approaches to account for errors in influence functions or related approaches. One approach would be to minimize the *monetary error* associated with i*nfluence estimation error, i.e.* the difference between a dividend calculated via true LOO loss vs. a dividend calculated via influence estimates. *Implementers* could do so by restricting how unequal dividends can become. For instance, under the *Binning* approach we used (above-median gets 2x), if influence functions underestimate the influence of a very helpful observation by a factor of more than ½, the corresponding dividend can at most be reduced by ½. In general, an approach that makes dividends more equal (e.g. above-median gets 1.1x) will further reduce the impact of error.

**Demographic Disparities**
Our results suggest that demographic disparities can realistically manifest in data dividends. Adding robustness to this critical high-level result, Ghorbani and Zhou also identified demographic disparities in their data valuation experiments with patient readmission data [22].

Under meritocratic approaches to *Observation Valuation*, a contributor's impact on a model will be dependent on the features used in model. In many cases features reflect descriptive attributes about a person and elements of a person's identity. Contributors may have no ability to modify these features (without lying), let alone interest in doing so. Thus, the notion of some data having more "merit" than other data becomes questionable. For such cases, "meritocratic" *Observation Valuation* may be poorly motivated and create counterproductive outcomes. Moreover, the literature on social computing has identified many online platforms with demographic contributions biases [11, 25, 28, 53]. Thus, using contributions patterns to assign payments can also exacerbate existing biases.

Work on machine learning fairness has highlighted that addressing the disparate impacts of intelligent technologies is extremely challenging [7], and our results suggest that doing so for data dividends may be similarly difficult. Regarding concerns around meritocratic *Observation Valuation*, it may be that techniques designed to increase the fairness of machine learning models could also improve the fairness of related data dividends (e.g. using fairness-aware models). However, our findings reinforce the need for caution and additional research along these lines.

**Nearly Uniform Observation Valuation**
Our results suggest that there is very little difference between using LOO loss with a *Shift* transform and assigning each observation the same value. Similarly, in studying the Shapley value approach to data valuation, Jia et al. provided a theorem stating that for all stable learning algorithms, LOO loss can be approximated with bounded error by a uniform distribution [35]. Furthermore, Jia et al. highlight that "a broad variety of learning algorithms are stable" [35]. This means there is a large class of technologies for which simple fixed amount per observation dividends are almost indistinguishable from estimating LOO loss change and applying the *Shift* transform

**LIMITATIONS**
As emphasized above, this paper was meant to help start the process of developing an empirically-grounded decision framework for data dividends. By no means does this paper finish this process. There are innumerable important directions of future work, both in terms of simulations and design space articulation.

With respect to simulations, we examined only straightforward transforms, used only a subset of ML approaches, and focused only on supervised classification (vs. regression, unsupervised learning, etc.) tasks. Valuable future work can be done expanding our research in all of these directions, in particular considering neural networks.

In terms of unexplored design dimensions, we did not consider in detail regions of the design space that relate to forward-looking data dividends and their potential for incentivizing data contributions. Future work should also consider how this region of the design space relates to the large literature on crowdwork marketplaces, although existing work in mechanism design provides insight into observation valuation for yet-to-be-contributed data, e.g. [1]. We do note that our findings related to data valuation have implications for future work on data markets – the potential for highly unequal distributions and demographic disparities likely apply to data markets that use meritocratic valuation. We also did not explore design choices that involve "data debt" because we did not believe them to be ecologically valid for early-stage data dividends. Finally, there may design dimensions that have yet to be uncovered.

Future work should also consider the effect of data dividends on national and global inequality, i.e. future data dividend simulations should be integrated with macroeconomic models. Even if the distribution of dividends for a given task is highly unequal, it is still possible to reduce income inequality at a high level, e.g. if the "superstars" in the highly-unequal data dividends we observed are people on the lower end of the income distribution.

**CONCLUSION**
Data dividends have been discussed – including by prominent politicians – as a way to mitigate inequality caused by increasingly intelligent technologies. In this paper, we provide an initial exploration of the high-stakes design space of data dividends. Using influence function techniques from the machine learning interpretability literature, we simulate data dividends in a variety of contexts. Our results suggest that the outcomes of data dividends can be unpredictable, with potentially harmful – rather than helpful – effects on inequality. Our results highlight the complexity of the data dividend design space and the importance of using simulations and careful design when implementing data dividends.